\documentclass[11pt]{article}
\usepackage{moriond,epsfig,graphics}

\def\be{\begin{equation}}
\def\ee{\end{equation}}
\def\bea{\begin{eqnarray}}
\def\eea{\end{eqnarray}}

\begin{document}
\vspace*{4cm}
\title{LATTICE QCD AT FINITE TEMPERATURE}

\author{ P. Petreczky }

\address{Department of Physics, Brookhaven National Laboratory,\\
Upton NY 11973, USA}

\maketitle\abstracts{
I review recent progress in lattice QCD at finite
temperature. Results on the transition temperature
will be summarized. Recent progress in understanding
in-medium modifications of interquark forces 
and quarkonia spectral functions at finite 
temperatures is discussed.
}

\section{Introduction}
\label{sec:intro}

It is expected that strongly interacting matter shows qualitatively
new behavior at temperatures and/or densities which are 
comparable or larger than the typical hadronic scale.
It has been argued that under such extreme conditions 
deconfinement of quarks and gluons should take place,
i.e. thermodynamics of strongly interacting matter could
be understood in terms of this elementary degrees of freedom
and this new form of matter was called 
{\em Quark Gluon Plasma}~\cite{shuryak_rev}. 
On the lattice the existence of the deconfinement  transition 
at finite temperature was first shown in
the strong coupling limit of QCD \cite{firstlat},
followed by numerical 
Monte-Carlo studies of lattice SU(2) gauge theory which confirmed it
\cite{firstlat1}. 
Since these pioneering studies QCD at finite temperature
became quite a large subfield of lattice QCD (for recent
reviews on the subject see Refs. \cite{edwin,katz,petreczkylat04}). 
One of the 
obvious reasons for this is 
that phase transitions can be studied only
non-perturbatively. But even at high temperatures the physics
is non-perturbative beyond the length scales $1/(g^2~T)$
($g^2(T)$ being the gauge coupling) \cite{linde}. Therefore lattice QCD
remains the only tool for theoretical understanding of the
properties of strongly interacting matter under extreme condition
which is important for the physics of the early universe as well as heavy ion 
collisions.

\section{Finite temperature transition in full QCD}

\begin{figure}
\centerline{
\psfig{file=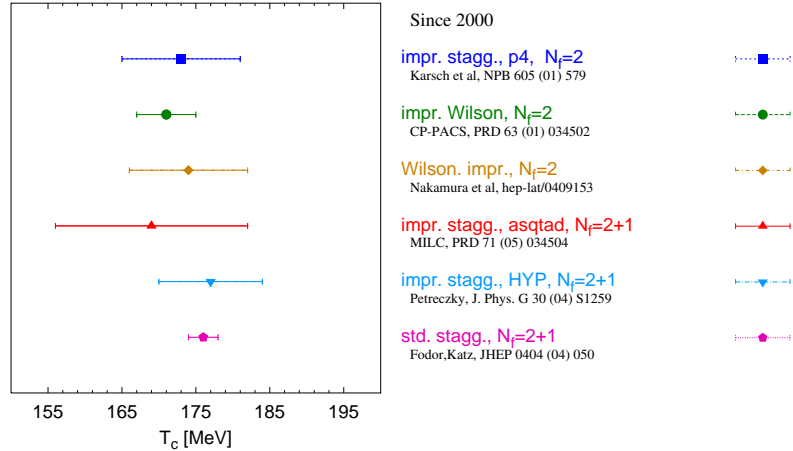,height=3.5in,angle=-90}
}
\caption{Summary of lattice results on the transition
temperature $T_c$ taken from Refs. 
\label{fig:tc}}
\end{figure}

One of the basic questions we are interested in is what is
the nature of the transition to the new state of mater and 
what is the temperature where it happens
\footnote{I will talk here about the QCD finite temperature transition
irrespective whether it is a true phase transition or a crossover
and $T_c$ will always refer to the corresponding temperature.}.
In the case of QCD without dynamical quarks, i.e. SU(3) gauge theory
these questions have been answered. It is well
established that the phase transition is 1st order \cite{fukugita89}. 
Using standard and improved actions the corresponding transition
temperature was estimated to be $T_c/\sqrt{\sigma}=0.632(2)$ 
\cite{edwin} ($\sigma$ is the string tension).
The situation for QCD with dynamical quarks is much more difficult.
Not only because the inclusion of dynamical quarks increases the 
computational costs by at least two orders of magnitude but also
because the transition is very sensitive to the quark masses.
Conventional lattice fermion formulations break global 
symmetries of continuum QCD (e.g. staggered fermion
violate the flavor symmetry) which also introduces 
additional complications. 
Current lattice calculations suggest
that transition in QCD for physical quark masses is not a
true phase transition but a crossover 
\cite{karsch01,frithjoflat03,fodor04,hyp,milcthermo}.
Recent lattice results for the transition temperature $T_c$ from
Wilson fermions \cite{cppacsnf2,nakamuralat04}, 
improved \cite{karsch01,hyp,milcthermo} and unimproved staggered fermions
\cite{fodor04} with 2 and 2+1 flavors of dynamical quarks 
are summarized in Fig. \ref{fig:tc}. The errors shown 
in Fig. \ref{fig:tc} are only statistical with the exception 
of the data point from the MILC collaboration, 
where the large error partly comes from the continuum 
extrapolation and also includes systematic error in scale setting.
Since the ``critical'' energy density $\epsilon_c=\epsilon(T_c)$ 
(i.e. the energy density at 
the transition) scales as $T_c^4$ the error in $T_c$ is the dominant source
of error in $\epsilon_c$ \cite{petreczkylat04}.   

\section{Heavy quarks at finite temperature}

\begin{figure}                                                                 
\centerline{                                                                   
\epsfig{file=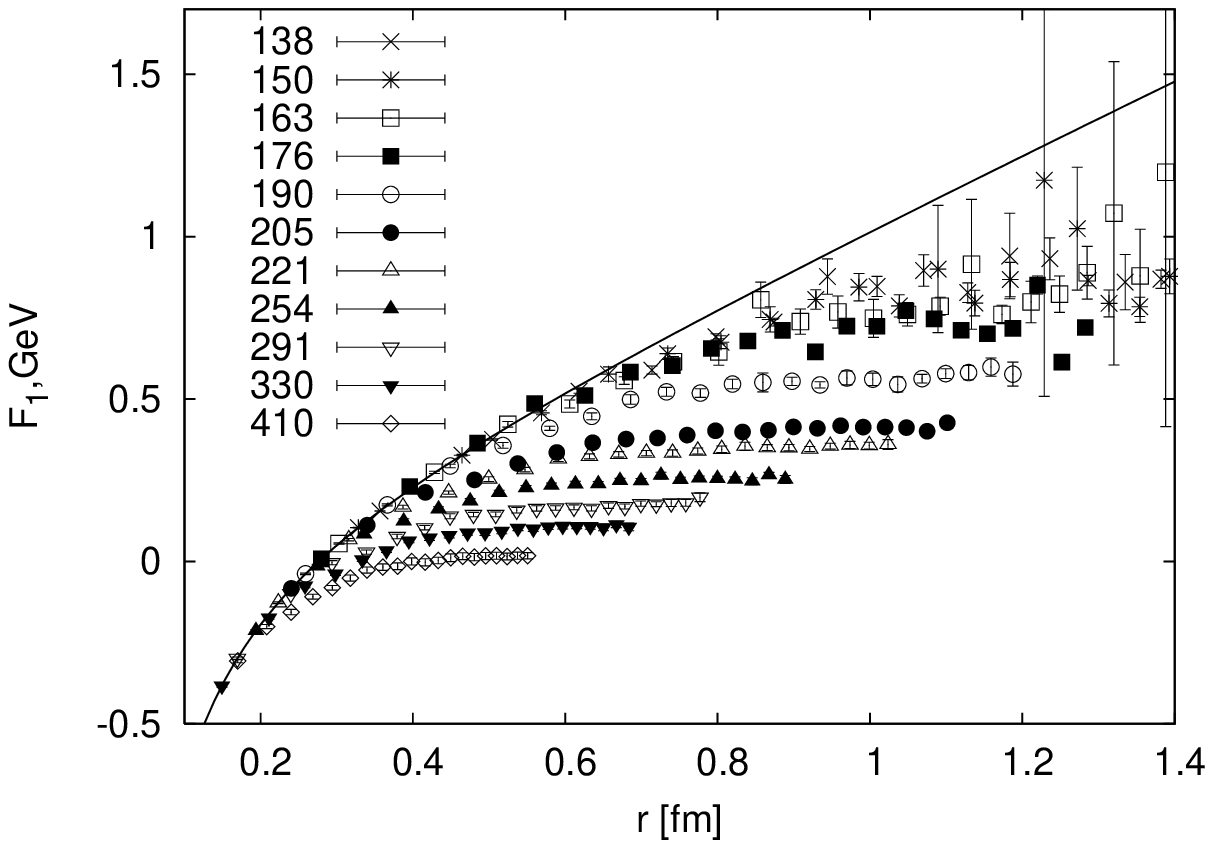, height=5cm}
\epsfig{file=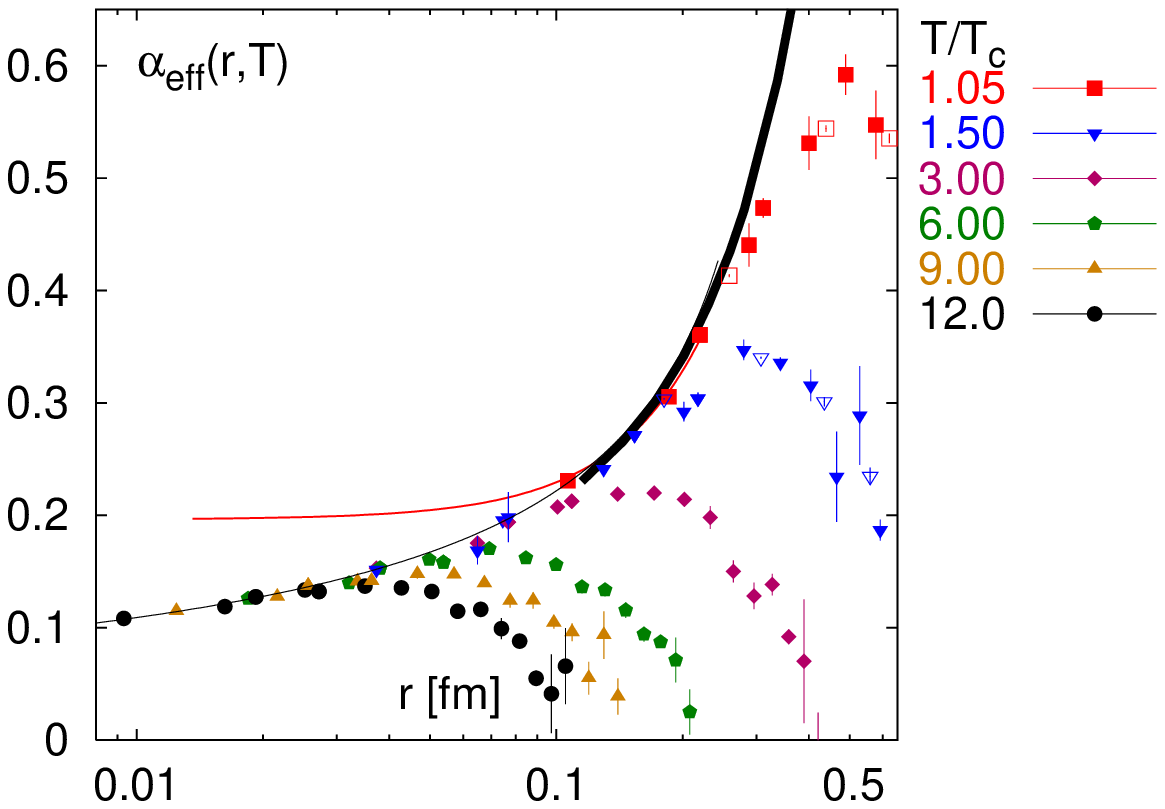, height=5cm}                       
}                                                                              
\caption{The singlet free energy in three flavor QCD at different
temperatures in MeV (left) and the coupling constant $\alpha_s$
at finite temperature (right).}        
\label{f1fig}                                                                
\end{figure}

In this section I am going to summarize some recent progress made
in understanding the interaction of heavy quarks at finite
temperature. Apart from being an interesting problem from a theoretical
perspective understanding the interaction of heavy quarks at finite 
temperature also is
very important for phenomenology. It has been suggested
that quarkonium suppression due to color screening at high 
temperatures can serve as signature of Quark Gluon Plasma formation
in heavy ion collisions \cite{MS86}.
For static quarks one can calculate the free energy difference
for the system with static quark anti-quark pair and the system
without static charges. This quantity is often referred to as 
finite temperature potential, though it
should be emphasized that it is a free energy and thus contains
an entropy contribution \cite{okacz02}. In Fig. \ref{f1fig} I show
the free energy of static $Q\bar Q$ in the singlet state calculated
in three flavor QCD \cite{petrov04}. As one can see from the Figure 
the free energy goes to a constant at distances $r>0.9$ fm at low temperatures.
This happens because once enough energy is accumulated the string can break
due to creation of a light quark-antiquark pair. As the temperature increases
the distance where the free energy levels off becomes temperature dependent
and decreases with increasing temperature. This reflects the onset of
chromo-electric screening. 
Similar results have been obtained in two flavor QCD 
\cite{okaczhard04,okacz05}.

\begin{figure}
\centerline{
\epsfig{file=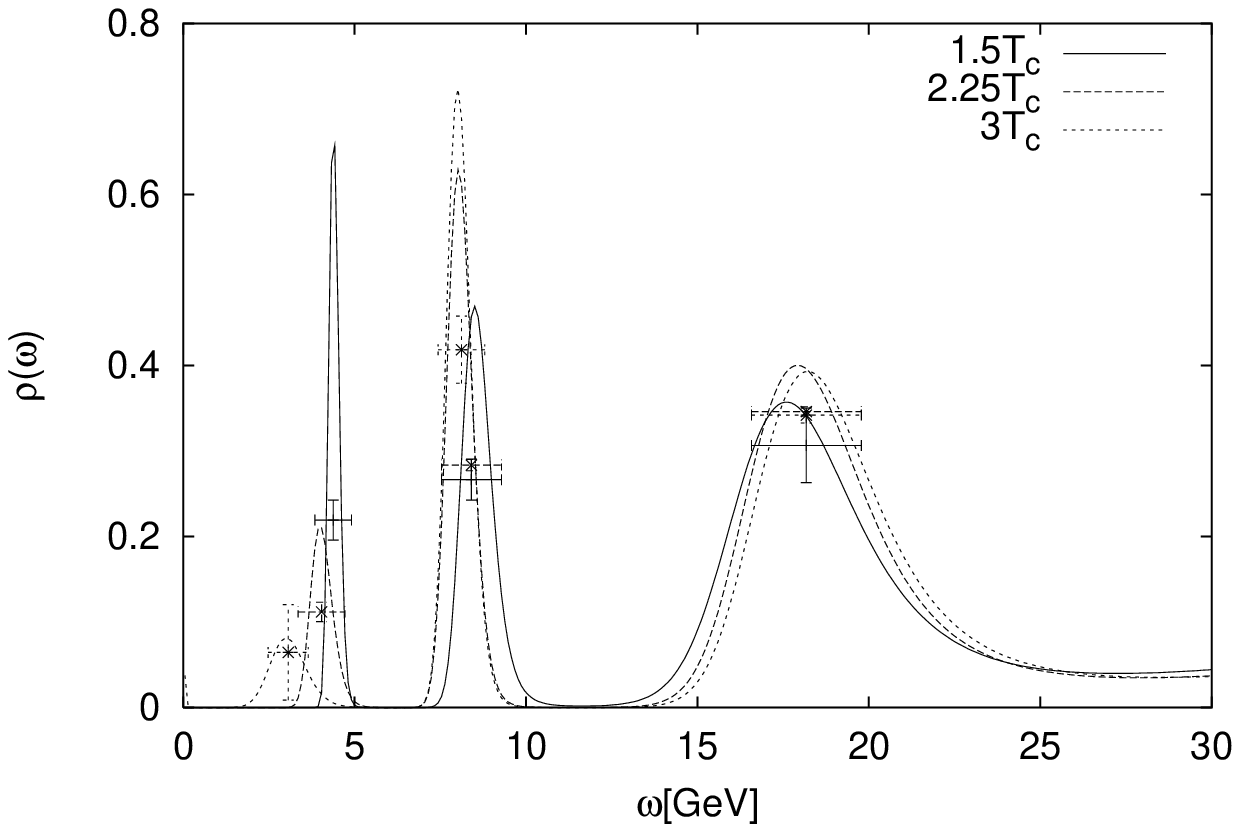, height=5cm}
\psfig{file=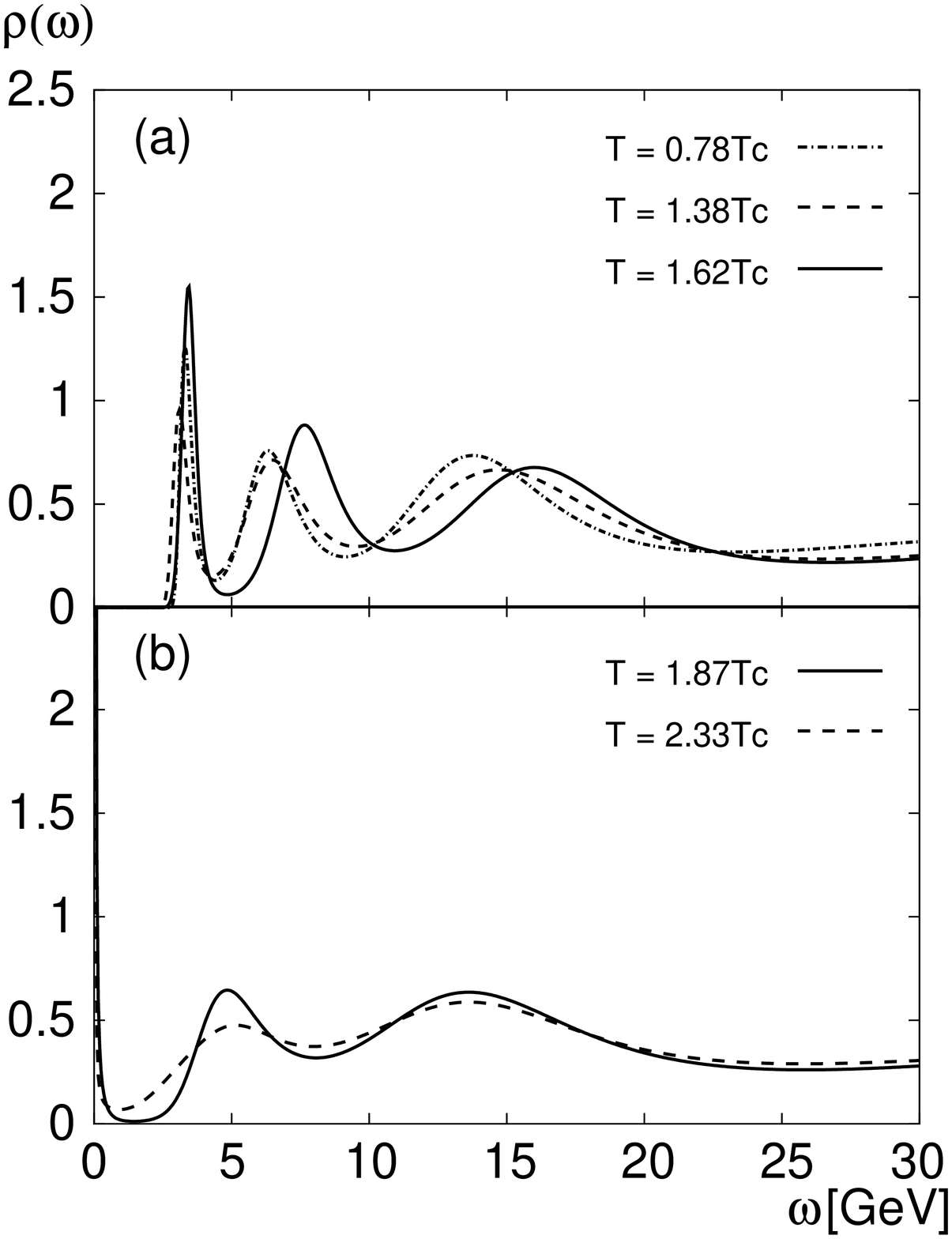, height=7cm}
}
\caption{The $J/\psi$ spectral functions from Datta et al.
(left) and from Asakawa and Hatsuda (right).}
\label{spf_fig}
\end{figure}
It should be noticed that
at short distances ($r < 0.4$ fm) 
the free energy of static $Q\bar Q$ is temperature
independent. As expected at short distance medium effects are not important.
This is also confirmed by studies of the coupling constant 
at finite temperature \cite{okacz04} which I also show in Fig. \ref{f1fig}.
The running of the coupling constant at finite temperature is controlled 
by the distance between the static  quarks and its value is never
larger than at zero temperature \cite{okacz04}.
This disfavors the picture of strongly coupled plasma where $\alpha_s$
runs to large value above the transition temperature \cite{shuryak03}.

Though the study of the free energy of a static quark anti-quark pair
gives some useful insight into the problem of quarkonium binding at high
temperatures ( for a recent review on this subject see Ref. \cite{petr05} ), 
it is not sufficient for detailed understanding quarkonium properties in this
regime. To gain quantitative information on this problem quarkonium 
correlators and spectral functions should be studied at finite 
temperature. Such studies became possible only recently and still are
restricted to the quenched approximation 
\cite{umeda02,asakawa04,datta04,dattaqm04,dattasewm04,petrovhard04}.
The results of these studies for charmonia 
are summarized in Fig. \ref{spf_fig}.
The $1S$ states ($J/\psi, \eta_c$) seem to survive 
to temperatures as high as $1.6T_c$
(maybe even higher, cf. the figure) while the 1P states 
($\chi_{c0},\chi_{c1}$) are dissolved at $1.1T_c$ \cite{datta04}.
The survival of the $1S$ state is also confirmed by 
Umeda et al \cite{umeda02}.
The temperature dependence of the charmonia correlators also
suggests that the properties of $1S$ charmonia are not affected 
significantly above $T_c$, at least at zero spatial momentum 
\cite{datta04,dattasewm04}. As for bottomonia only preliminary results
exists showing that $\Upsilon$ can exist in the plasma up to much
higher temperatures \cite{petrovhard04} but surprisingly enough the 
$\chi_b$ state is dissociated at temperatures smaller than $1.5T_c$ 
\cite{petrovhard04}.

\section*{Acknowledgments}
This work has been authored under the contract 
DE-AC02-98CH10886 with the U.S. Department of energy.
I would like to thank Frithjof Karsch for careful reading of the
manuscript.


\section*{References}

\begin{thebibliography}{99}

\bibitem{shuryak_rev}
  E.~V.~Shuryak,
  Phys.\ Rept.\  {\bf 61} (1980) 71.


\bibitem{firstlat}
A.M. Polyakov, Phys. Lett. {\bf B72} (1978) 477; 
L. Susskid, Phys. Rev D {\bf 20} (1979) 2610

\bibitem{firstlat1}
L.D. McLerran and B. Svetistky, Phys. Lett. B {\bf 98} (1981) 195;
J. Kuti, J. Pol\'onyi and K. Szlach\'anyi, Phys. Lett. B{\bf 98} (1981) 199;
J. Engels et al, Phys. Lett. B {\bf 101} (1981) 89

\bibitem{edwin}
E.~Laermann and O.~Philipsen,
Ann.\ Rev.\ Nucl.\ Part.\ Sci.\  {\bf 53} (2003) 163

\bibitem{katz}
S.D. Katz, Nucl.\ Phys.\ Proc.\ Suppl.\  {\bf 129} (2004) 60

\bibitem{petreczkylat04}
P.~Petreczky,
  Nucl.\ Phys.\ Proc.\ Suppl.\  {\bf 140} (2005) 78

\bibitem{linde}
A.D. Linde, Phys. Lett. B {\bf 96} (1980) 289

\bibitem{fukugita89}
M. Fukugita, M. Okawa, A. Ukawa, Phys. Rev. Lett. {\bf 63} (1989) 1768

\bibitem{karsch01}
F. Karsch, E. Laermann, A. Peikert, Nucl. Phys. B {\bf 605} (2001) 579

\bibitem{frithjoflat03}
F. Karsch et al., Nucl. Phys. B (Proc. Suppl.) {\bf 129-130} (2004) 614

\bibitem{fodor04}
Z.~Fodor and S.~D.~Katz,
JHEP {\bf 0404}, 050 (2004)

\bibitem{hyp}
P.~Petreczky,
  J.\ Phys.\ G {\bf 30} (2004) S1259.

\bibitem{milcthermo}
  C.~Bernard {\it et al.}  [MILC Collaboration],
  Phys.\ Rev.\ D {\bf 71} (2005) 034504

\bibitem{cppacsnf2}
A.~Ali Khan {\it et al.}  [CP-PACS Collaboration],
Phys.\ Rev.\ D {\bf 63} (2001) 034502

\bibitem{nakamuralat04}
Y. Nakamura et al, hep-lat/0409153

\bibitem{MS86}
T.~Matsui and H.~Satz,
  Phys.\ Lett.\ B {\bf 178} (1986) 416.

\bibitem{okacz02}

 O.~Kaczmarek, F.~Karsch, P.~Petreczky and F.~Zantow,
  Phys.\ Lett.\ B {\bf 543} (2002) 41

\bibitem{petrov04}
 P.~Petreczky and K.~Petrov,
  Phys.\ Rev.\ D {\bf 70} (2004) 054503
  [arXiv:hep-lat/0405009].

\bibitem{okaczhard04}
 O.~Kaczmarek and F.~Zantow,
  arXiv:hep-lat/0502011.



\bibitem{okacz05}
  O.~Kaczmarek and F.~Zantow,
  arXiv:hep-lat/0503017.


\bibitem{okacz04}
 O.~Kaczmarek, F.~Karsch, F.~Zantow and P.~Petreczky,
  Phys.\ Rev.\ D {\bf 70} (2004) 074505
  [arXiv:hep-lat/0406036].

\bibitem{shuryak03}

  E.~V.~Shuryak and I.~Zahed,
  Phys.\ Rev.\ C {\bf 70} (2004) 021901

\bibitem{petr05}
 P.~Petreczky,
  arXiv:hep-lat/0502008.

\bibitem{umeda02}

  T.~Umeda, K.~Nomura and H.~Matsufuru,
  Eur.\ Phys.\ J.\ C {\bf 39S1} (2005) 9


\bibitem{asakawa04}

  M.~Asakawa and T.~Hatsuda,
  Phys.\ Rev.\ Lett.\  {\bf 92} (2004) 012001

\bibitem{datta04}
 S.~Datta, F.~Karsch, P.~Petreczky and I.~Wetzorke,
  Phys.\ Rev.\ D {\bf 69} (2004) 094507

\bibitem{dattaqm04}
  S.~Datta, F.~Karsch, P.~Petreczky and I.~Wetzorke,
  J.\ Phys.\ G {\bf 30} (2004) S1347

\bibitem{dattasewm04}
  S.~Datta, F.~Karsch, S.~Wissel, P.~Petreczky and I.~Wetzorke,
  arXiv:hep-lat/0409147.

\bibitem{petrovhard04}

  K.~Petrov,
  arXiv:hep-lat/0503002.

\end{thebibliography}
\end{document}